\documentclass[12pt]{article}
\usepackage{graphicx}
\title{Remarks on homo- and hetero-polymeric aspects of protein
folding}
\author{T. Garel \footnote{Member of CNRS}\\
Service de Physique Th\'eorique, CEA/DSM/SPhT\\
Unit\'e de recherche associ\'ee au CNRS\\
91191 Gif sur Yvette Cedex, France\\}

\def\be{\begin{equation}}
\def\bea{\begin{eqnarray}}
\def\ee{\end{equation}} 
\def\eea{\end{eqnarray}} 
\begin{document}
\maketitle
\vskip 4mm

\begin{abstract}

Different aspects of protein folding are illustrated
by simplified polymer models. Stressing the diversity of side chains
(residues) leads one to view folding as the freezing 
transition of a heteropolymer. Technically, the most common approach
to diversity is randomness, which is usually implemented in two body
interactions (charges, polar character,..). On the other hand, the
(almost) universal character of the protein backbone suggests that
folding may also be viewed as the crystallization transition of an
homopolymeric chain, the main ingredients of which are the peptide
bond and chirality (proline and glycine notwithstanding). The model of 
a chiral dipolar chain leads to a unified picture of secondary
structures, and to a possible connection of protein structures with
ferroelectric domain theory.

\end{abstract}
\vskip 4mm
{\bf Lectures given at the Trieste Workshop on Proteomics: Protein
Structure, Function and Interaction, May 5-16 2003}
\vskip 4mm
\hfill\mbox{Saclay T03/051}\\ \noindent \mbox{ }\\
\newpage

\section{Introduction}
\label{intro}

Proteins are polymers, which have the property of folding
reversibly into a single geometrical shape with biological
activity. The folding process can be modeled as a
phase transition from a high temperature coil phase to a low
temperature compact phase (other parameters, e.g. pH, may also
trigger the transition). Both phases require a very detailed
description of the monomers, not to mention the chemistry of water.
These notes aim at presenting some of the theoretical
approaches to the folding transition. The interested reader
can find more details in the short list of references at the end of
the paper. 

Section \ref{Chemistry} gives a brief description of the twenty
monomers (amino acids) which are the building blocks of proteins. A
protein of $N$ amino acids can be characterized by its primary
structure, i.e. by specifying which amino acid is actually at position
$({\rm i})$ along the chain (with ${\rm i}=1,2...N$ and $N$ of the
order of a few hundreds). 

At first sight, a protein may be considered as a heteropolymer, with
amino acid $({\rm i})$ being characterized by its electric charge
$(q_{i})$, its hydrophilicity $(\lambda_{i})$.... A 
closer look at a protein chain reveals that this heteropolymer is made
of an almost periodic backbone and of different side chains
(residues). This almost periodic backbone is almost protein
independent, leading to a homopolymeric approach to the folding
transition. This homopolymeric view is supported by the ubiquitous
existence of helices and sheets in proteins (which are called the
secondary structures). Folded proteins illustrate the coexistence of
specific features (primary sequence, type of biological activity,...)
and universal features \cite{Bran_Too, Crei1} such as helices and
sheets (one may also speculate about the universal character of other
issues, such as the very existence of a biological activity or the
aggregated structures in amyloid-like diseases \cite{Guo}). 

Some of these properties will be illustrated, mostly in a pictorial
way, on a small protein (1aps, $N=98$) in Section \ref{aps}. This
example shows that one is not concerned by the theorists'
thermodynamic limit: proteins have a finite size, pointing towards an
important role of the surface, and therefore of the solvent
(i.e. water for globular proteins). More numbers pertaining to the
folding process will be given in Section \ref{nombres}.

I think it is fair to say that the homopolymeric aspects of proteins
have been so far less studied. Since there are numerous reviews on
their heteropolymeric properties, I will mostly deal here with
homopolymeric models, and conclude on the interplay between both types 
of properties.

\section{Chemistry \cite{Bran_Too, Crei1}} 
\label{Chemistry}

\begin{itemize}

\item A brief description of the monomers 

\begin{itemize}
\item  There are twenty different types of monomers
\vskip 3mm
ASP, GLU, LYS, ARG, ALA, VAL, PHE, {\bf PRO}, MET, ILE, LEU, SER, THR,
TYR, HIS, {\bf CYS}, ASN, GLN, TRP, {\bf GLY} 

\item These monomers are amino acids (exception: {\bf PRO})
\vskip 3mm
\centerline{${\rm {{\bf H_{2}N}-C_{\alpha}HR-{\bf COOH}}}$ }

e.g.  

\centerline{ALA:   ${\rm R= CH_3}$}

\vskip 3mm

\centerline{GLU:   ${\rm R= (CH_{2})_{2}-COO^{-}}$}

\item The amino acids are chiral (exception: {\bf GLY})
\vskip 3mm
\centerline{${\rm {H_{2}N-{\bf C_{\alpha}}HR-COOH}}$ }
\vskip 3mm
Sitting on the ${\rm C_{\alpha}-H}$ bond and looking towards the
${\rm C_{\alpha}}$ atom, one sees the ${\rm CO-R-N}$ sequence in a
clockwise way. 

\item Residues ${\rm R}$ have different properties
\vskip 3mm
\centerline{${\rm {H_{2}N-C_{\alpha}H{\bf R}-COOH}}$ }

\begin{itemize}
\item Charged residues  ASP (-), GLU (-), LYS (+), ARG (+)
\item Polar residues TYR, HIS, ASN, GLN
\item Rather polar residues {\bf PRO}, THR, ALA, {\bf GLY}, SER
\item Hydrophobic residues VAL, PHE, MET, ILE, LEU, TRP, {\bf
CYS}

\end{itemize}

\end{itemize}

\item From monomer to polymer

\begin{itemize}

\item Formation of the peptide bond

\begin{figure}
\begin{center}
\includegraphics[height=4cm]{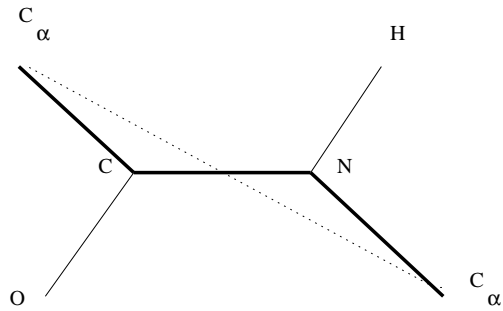}
\caption{Peptide bond: the ${\rm C}_{\alpha}-{\rm C}_{\alpha}$ virtual
bond is roughly perpendicular to the CO and NH bonds.} 
\end{center}
\end{figure}

Geometrical constraints (stemming from quantum chemistry), imply that
consecutive ${\rm {{\bf C_{\alpha}}-CO-NH-{\bf C_{\alpha}}}}$ atoms
are coplanar, with the C-O (and N-H) bonds being roughly perpendicular
to the virtual ${\rm {{\bf C_{\alpha}}-{\bf C_{\alpha}}}}$ bond.

At long distance, the charge distribution on the peptide bond is
dominated, in a first approximation, by an electric dipole parallel to
${\rm {OC}}$ and of order four Debyes. At shorter distances, this
charge distribution gives rise to hydrogen bonds. 

With the exception of residues GLY and PRO, the backbone chain can be
written 
\vskip 3mm
${\rm {...NH-{\bf C_{\alpha}}-CO-NH-{\bf C_{\alpha}}-CO-NH-{\bf
C_{\alpha}}-CO-NH-{\bf C_{\alpha}}-CO...}}$ 
\vskip 3mm
\newpage

\item Chirality of the main chain

The chirality of the amino acids, together with the steric constraints
gives to the main backbone chain some sort of chirality. More
precisely, taking four atoms 1234 along the main backbone chain, the
associated dihedral angle $\widehat {1234}$ is defined as the angle of
the (123) and (234) planes. It is shown in Figure 2, with the 2-3 bond
coming out of the paper. The convention is such that the above dihedral
angle $\widehat {1234}$ is negative. Note that $\widehat
{1234}=\widehat {4321}$.

\begin{figure}
\begin{center}
\includegraphics[height=4cm]{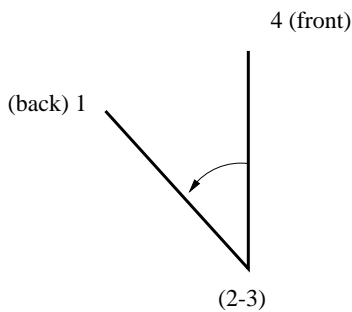}
\caption{Defining the dihedral angle $\widehat {1234}$ for atoms 1,2,3,4}
\end{center}
\end{figure}

The chirality of the chain is associated with the fact that positive
and negative dihedral angles do not have the same steric constraints
and therefore not the same (free) energies. In 
particular, Ramachandran's plots for a given residue, are given for
$\phi=\widehat {1234}$ (resp. $\psi$) where $1234={\rm CNC_{\alpha}C}$
(resp. $1234={\rm NC_{\alpha}CN}$). Except for GLY, these plots show that
helices and sheets correspond to rather well defined (and non
symmetric) regions in $\phi,\psi$ space. 

One may also define Ramachandran's plots for side chains, but we will
neglect their role in this paper.

\end{itemize}

\vskip 3mm

So with the (important) exceptions of residues GLY and PRO, the main
backbone chain can be represented in a homopolymeric way
(see e.g. \cite{Head}). One may then consider the chiral
(hydrogen-bonding+dipolar) chain as a good description of secondary
structures in proteins. 

\end{itemize}

\section {Example}
\label{aps}

I will illustrate some of the  previous points on a specific protein
(1aps).

\begin{itemize}

\item Primary structure ( $N=98$ residues)

 (i=1) SER-THR-ALA-ARG-PRO-LEU-LYS-SER-VAL-ASP-TYR-GLU-VAL-

 -PHE-GLY-ARG-VAL-GLN-GLY-VAL-CYS-PHE-ARG-MET-TYR-ALA-

 -GLU-ASP-GLU-ALA-ARG-LYS-ILE-GLY-VAL-VAL-GLY-TRP-VAL-

 -LYS-ASN-THR-SER-LYS-GLY-THR-VAL-THR-GLY-GLN-VAL-GLN-

 -GLY-PRO-GLU-GLU-LYS-VAL-ASN-SER-MET-LYS-SER-TRP-LEU-

 -SER-LYS-VAL-GLY-SER-PRO-SER-SER-ARG-ILE-ASP-ARG-THR-

 -ASN-PHE-SER-ASN-GLU-LYS-THR-ILE-SER-LYS-LEU-GLU-TYR-

 -SER-ASN-PHE-SER-VAL-ARG-TYR (i=98)

\begin{figure}
\begin{center}
\includegraphics[height=7cm]{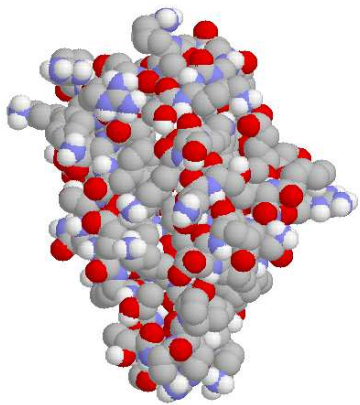}
\caption{Native spacefilled structure of 1aps}
\label{figure2}
\end{center}
\end{figure}

\begin{figure}
\begin{center}
\includegraphics[height=7cm]{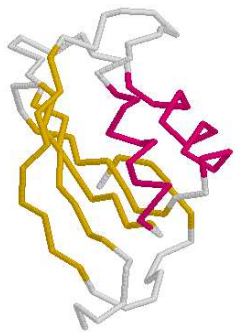}
\caption{Backbone of 1aps: Helices 1 (PHE 22-ILE 33) and 2 (GLU 55-LEU
65). Sheet ( strand 1 (LYS 7-VAL 13); strand 2 (VAL36-THR 42); strand
3 (THR 46-GLY 53); strand 4 (ARG 77-THR 85); strand 5 (ASN 93-ARG 97))}
\label{figure1}
\end{center}
\end{figure}

\newpage
\item Compactness of the folded structure (Figure 3)

The total number of atoms is of order one thousand, linking
protein folding with cluster physics (with a chain constraint)
\cite{Berry,Karplus}.

\item Charged residues (Figure 5)

There are 26 charged residues (three ASP, seven GLU, seven ARG and
nine LYS). For electrostatic reasons, they are located on the protein
surface. More generally, since there are four (out of twenty) charged
residues in usual conditions, and assuming equipartition, a protein
of N residues has N/5 charged residues to be placed on the surface
(which scales likes $N^{2 \over 3}$), leading to an estimate $N \sim
125$ for a typical single domain protein.

\item The main backbone chain (Figures 4, 6, 7)

As mentioned above, the main chain is homopolymeric with the
exception of three PRO and eight GLY residues. This homopolymeric
character is illustrated by the CO bonds. The (roughly) ferroelectric
order of helices and the (roughly) antiferroelectric character of
sheets result from hydrogen bonding and dipolar-like forces.

\begin{figure}
\begin{center}
\includegraphics[height=7cm]{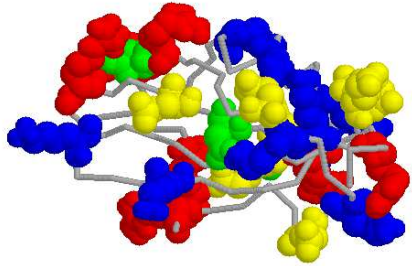}
\caption{Charged residues of 1aps}
\label{figure 4}
\end{center}
\end{figure}

\end{itemize}

\begin{figure}
\begin{center}
\includegraphics[height=7cm]{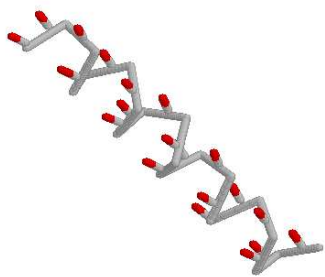}
\caption{Homopolymeric backbone: CO bonds of a helix}
\end{center}
\end{figure}

\begin{figure}
\begin{center}
\includegraphics[height=7cm]{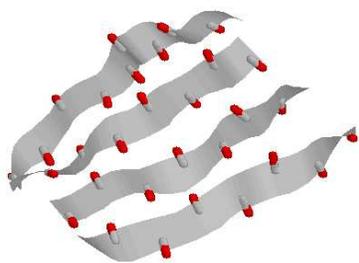}
\caption{Homopolymeric backbone: CO bonds of a sheet}
\end{center}
\end{figure}
\newpage

I could not find the active site of 1aps on the web. From a
physical point of view, one would like to understand how the primary
sequence somehow encodes the native structure (to further extract the active
site from the native structure is not easy).

\section {Numbers \cite{Bran_Too, Crei1}}
\label{nombres}
\vskip 3mm
\begin{itemize} \itemsep 10pt    

\item Energy scales

At room temperature ${\rm T_0}$, the equivalence between chemical and
physical units is  ${\rm {k_{B} T_0}} \sim 0.6 \ {\rm {kcal/mole}}
\sim {1 \over 40} \ {\rm {eV/part}}$. A hydrogen bond has an energy of 
order 2-6 kcal/mole. A covalent bond has an energy of order $50-200$ 
kcal/mole. In the folded state, two consecutive peptide bonds have an
energy of order 1-3 kcal/mole ( obtained from ${p_0^2 \over 4 \pi
\epsilon_0 \epsilon_r d^3}$, with $p_0 \sim 4$ Debyes, $d \sim 4$ \AA,
and where $\epsilon_r$ is believed to be of order $2-5$). Finally, van
der Waals attraction energies are of order 0.3-1 kcal/mole.

The folding transition is first order, with an
entropy loss of order $k_{B}$ or less per residue (possibly
raising questions on the applicability of Classical Statistical
Mechanics).

The dynamics of the folding is governed by energy barriers:
the range of folding times is of order $10^{-3}-1s$, and even
longer; it should be compared with microscopic times of order
$10^{-13}-10^{-15} s$. This suggests that the phase space of a protein
may have many trapping local minima, implying problems in numerical
simulations.

\item Geometry and Energy

A typical protein has something like $N \sim 100-500$ residues
(the number of atoms being of order a few thousands). Something like
half of the residues belong to the surface. A  typical linear size of
the folded molecule is $R \sim 50$ \AA. As can be seen from the
examples of 1aps and other proteins, the typical length of a helix
is 10-20 residues, that of a strand being smaller (may be 5-8
residues). In broad terms, proteins are clusters-with-a
chain-constraint (and a solvent). Geometrical constraints (bond
lengths, valence angles, van der Waals radii, chirality,..) play an
important role in proteins. The all-atom CHARMM energy
$E$ commonly used in numerical situations is given by \cite{Karplus2}

$$E =  \sum^{ }_{ {\rm bonds}}k_b
\left(b-b_0 \right)^2+ \sum^{ }_{ {\rm angles}}k_\theta \left(\theta -\theta_
0 \right)^2
$$
$$
+ \sum^{ }_{ {\rm dihedrals}}k_\phi( 1+ {\rm cos}(n\phi -\delta)) +
 \sum^{ }_{ {\rm impropers}}k_\nu
\left(\nu -\nu_ 0 \right)^2 
$$
$$
+ \sum^{ \ \ \ \ \ \prime}_{ i<j}4\varepsilon_{ ij}
\left\{ \left({\sigma_{ ij} \over r_{ij}} \right)^{12}- \left({\sigma_{ ij}
\over r_{ij}} \right)^6 \right\} 
 + \sum^{ \ \ \ \ \prime}_{ i<j}{332 \over
\varepsilon_ r} {q_iq_j \over r_{ij}} 
$$
\be
\label{Charmm}
\ee

where distances are in Angstr\"oms, angles in radians and $ E $ in $ {\rm
kcal/mol}$. Studying the dielectric permittivity $\varepsilon_ r$ is a 
difficult task; it is believed to be of order $2-5$ in a folded
protein. 

\item (Bio)chemistry

Before considering simplified models, let me recall a few facts about
real proteins. One should be cautious about in vivo vs in vitro
folding (role of chaperone molecules). Real proteins are not random
polymers, but have been evolution selected. Since (quantum) hydrogen
bonds are important in proteins, the use of Classical Statistical
Mechanics may be questioned. To this discouraging list, one may add
that water (for globular proteins) is a complicated, strongly
structured solvent. We will not even mention the biochemical activity
(e.g. the recognition of the active site by a ligand). The distance
between real proteins and simplified models is not to be underestimated.
 
\end{itemize}

\section{Simplified models: homopolymeric approach}

\subsection{Modeling hydrogen bonds in a compact phase}
\label{hbonds}

The discovery of helices and sheets by Pauling and Corey
\cite{Pauling_Corey1,Pauling_Corey2} relies on the
existence of short range hydrogen bonds. Let us consider first
an all-helix folded protein. What we have here is a competition
between local and global orders: the former favors helical (i.e. one
dimensional) structures (figure 8), whereas the latter favors compact
(i.e. three dimensional) structures.

This competition can be modeled as follows. We consider a polymer
chain on a cubic lattice, where the monomers interact with an
attractive van der Waals energy $\varepsilon_v$ (to ensure compactness 
at low temperature), and a curvature energy $\varepsilon_h,$ which
favors the alignment of two consecutive monomers. In this model, a
monomer represents a helical turn of the protein (two consecutive
helical turns allowing for the presence of hydrogen bonds).

The phase diagram depends on the ratio
$({\varepsilon_v \over \varepsilon_h})$; for simplicity, we will
restrict the discussion to fully compact structures 
($\varepsilon_v=\infty$). The case $\varepsilon_h=0$ is of interest in
the physics of hydrophobic chains at temperature below the $\theta$
point, which are known to possess a large entropy in the compact phase.
We will first study this model, and then consider the influence of the 
curvature term $\varepsilon_h$. Technical details are postponed to an
Appendix. 

\begin{figure}
\begin{center}
\includegraphics[height=3cm]{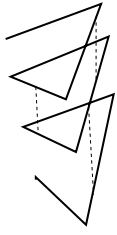}
\caption{Hydrogen bonds in a helix}
\end{center}
\end{figure}

\subsubsection{Entropy of a hydrophobic chain \cite{Orland}}
\vskip 3mm
We first define a Hamiltonian Path (HP) as a fully compact self
avoiding walk (SAW). The number of Hamiltonian Paths (HP) on the
lattice is formally described by

$${\cal N}_N=\sum_{(HP)}{\bf 1}$$

Introducing an n-component field $\vec {\varphi}_{\vec r}$ at each
point $\vec r$ of the lattice, and using the properties of the limit
$n \to 0$, it is shown in the Appendix that 
\be
\label{NHP1}
{\cal N}_N={\rm lim}_{{\rm n}\to 0} {1 \over {\rm n}} \int
{\cal D}{\vec {\varphi}_r} \ e^{-{1 \over 2}\sum_{(r,r')}{\vec 
{\varphi}}_{r} \ (\Delta^{-1})_{{\vec r}{\vec r'}} \ {\vec
{\varphi}}_{r'}} \prod_{r}\left({1 \over 2}{\vec
{\varphi}_r}^2\right)
\ee

where $\Delta_{{\vec r}{\vec r'}} =1 $ if $\vec r$ and $\vec r'$ are
nearest neighbour sites, and $0$ otherwise.

A homogeneous and isotropic saddle point evaluation $\vec
{\varphi}_{\vec r}=\vec {\varphi}$ yields  
\be
\label{Ntheta}
{\cal N}_N=\left({q \over e}\right)^N
\ee
with $q=2d=6$.

This calculation suggests that a collapsed homopolymeric
chain is a mixture of an exponentially large number of conformations,
and is therefore not able to describe the single conformation of a
native protein. For this reason, the folding transition is commonly
thought to be quite different from the $\theta$ transition.   
Note however that equation (\ref{Ntheta}) depends on the 
use of a homogeneous saddle point in equation (\ref{NHP1}), which is
consistent with the use of periodic boundary conditions. More general
boundary conditions (corresponding to non-homogeneous $\vec
{\varphi}_{\vec r}$) would give a expression of the form 
\be
\label{Ntheta2}
{\cal N}_N \sim A\mu^{N} \ {\widetilde {\mu}}^{N^{2/3}}
\ee
where $A$ is a constant, $\mu$ and $\widetilde \mu$ being respectively
bulk and surface connectivities.

\subsubsection{``Helices'' in a compact phase}
We now implement the competition between one- and three-dimensional
structures, by introducing a curvature energy 
$\varepsilon_h$. As mentioned above, this curvature energy favors
aligned consecutive ``monomers''  (or disfavors corners in the
HP). Denoting by $N_{corners}(HP)$ the number of corners in a given
(HP), the number of weighted (HP) is given by 

\vskip 2mm
$${\cal N}_h=\sum_{(HP, corners)} e^{-\beta N_{corners}(HP)\varepsilon_h}$$
where the summation runs over all possible (HP)'s and over
all possible corners, and where $\beta={1 \over T}$ is the inverse
temperature.  

Introducing $d$  ${\rm n}$-dimensional fields: ${\vec
{\varphi}}_{\alpha r}$ , $\alpha=1,2,...d$ for each lattice site
$(r)$, it is shown in the Appendix that
\be
\label{NHP2}
{\cal N}_h={\rm lim}_{{\rm n}\to 0} {1 \over {\rm n}} \int {\cal
D}{\vec 
{\varphi}}_{\alpha r} \ e^{-{1 \over 2}\sum_{\alpha=1}^{d} \sum_{(r,r')}{\vec 
{\varphi}}_{\alpha r} \ (\Delta_{\alpha}^{-1})_{{\vec r}{\vec r'}} \ {\vec
{\varphi}}_{\alpha r'}}
\prod_{r}\left({1 \over 2}\sum_{\alpha=1}^{d}{\vec
{\varphi}}_{\alpha r}^2+ e^{-\beta \varepsilon_h}
\sum_{\alpha<\gamma}{\vec {\varphi}}_{\alpha r} \cdot {\vec
{\varphi}}_{\gamma r} \right)
\ee

where $ \Delta^ \alpha_{\vec  r\vec  r^{\ \prime}} $ is 1 if $ \vec  r
$ and $ \vec  r^{\ \prime} $ are nearest neighbours in direction $
\alpha $  and $ 0 $ otherwise.

 Performing a homogeneous and isotropic saddle point in equation
(\ref{NHP2}) we get

\be
\label{HEL}
{\cal N}_h=\left({q(\beta) \over e} \right)^N
\ee

with an effective coordination number, $q(\beta)=2+2(d-1) e^{-\beta
\varepsilon_h}$.

A first order crystallization transition, describing the competition
between the entropy gain of making turns, and the corresponding energy
loss, occurs for $q(\beta_c)=e$. For $d=3$, the transition temperature
is $k_{\rm B} T_c = 0.58\ \varepsilon_h$. Below the transition, the
entropy is not extensive.
The average length of a helix is given by
$$ \ell  = {N\varepsilon _h\over U \left(\beta_ c \right)} $$
where $ U(\beta)  = - {\partial \over \partial \beta}  {\rm log} \
{\cal N}_h $ is the 
internal energy. Just above the transition, in $ d=3, $ the average
helix length is equal to $ \ell_ c = 3.78$, and is of ${\cal O} (\ell
=N^{1/3}) $ in the low temperature phase. 
Note that in this very simplified picture $ \ell_ c $ corresponds to a 
typical number of residues of the order of 15, since one monomer
corresponds to a helical turn, that is 3.6 residues. As seen from the
example of 1aps, and from numerical calculations using Hamiltonians such as
(equation(\ref{Charmm})), this is indeed the typical length of
$\alpha$-helices in proteins.

These results result from a homogeneous saddle point assumption
(implying the use of periodic boundary conditions). In a more correct
treatment, we expect a non extensive surface entropy of order $N^{2
\over 3}$ in the crystalline phase (corresponding to the fact that the
corners are on the surface of  the lattice). The  influence of boundary
conditions on the counting of (HP), with or without curvature 
energy, is a rather difficult subject \cite{Malakis}.

Finally, relaxing the constraint ($\varepsilon_v=\infty$), yields a phase
diagram where one may reach the crystallized phase either through a
$\theta$ transition followed by a second (liquid globule-crystal)
transition, or through a unique discontinuous coil-crystal
transition \cite{Do_Ga_Or}. One may thus consider that there are
two coil ``phases'', one above the $\theta$ transition, and the other
above the crystallization transition, differing by their short range
order. 

\subsubsection{Fully compact ``sheets''}

The extension to sheets (see figure 9) can be done with a slight
generalisation of the Hamiltonian Path formalism: a node of a path is
now to be interpreted as an amino acid.

\begin{figure}
\begin{center}
\includegraphics[height=3cm]{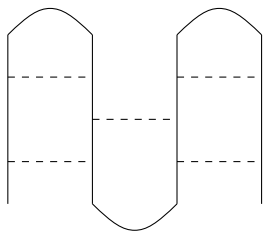}
\caption{Hydrogen bonds in a sheet}
\end{center}
\end{figure}

The model can be described in the following way.
Consider  a Hamiltonian path. To mimic the formation of $ {\rm CO-HN},
$ i.e. a hydrogen bond (H-bond), in $ \beta $-sheets, we 
allow an $ {\rm H} $-bond (energy gain $ \varepsilon_s ), $ whenever two pairs
of aligned links belong to two (non intersecting) neighbouring
strands. We do not make any distinction between parallel and
antiparallel sheets. Following the representation of the Appendix, and
performing an isotropic homogeneous saddle point, one also gets a
first order crystallization transition. The physics is very similar to
the cas of helices. Typical lengths of ordered strands are however
more difficult to estimate \cite{Bascle}.

\subsubsection{Conclusion on Hamiltonian Paths}

We have presented a simple model of the formation of secondary
structures in a dense phase, which is linked to polymer melting
theory. Starting from the coil state, one can reach the ``compact
state with secondary structures'' either directly or through other
compact phase(s). These results rest on a homogeneous saddle point
approximation, which corresponds to periodic boundary conditions.
The ``$n \to 0$'' approach \cite{PGG,Sarma} represents the chain via a
field $\vec 
{\varphi}_{r}$ and is therefore not appropriate to the description of
heteropolymeric properties ( where the information depends of the
curvilinear abcissa $(i)$ along the chain). In our presentation,
helices and sheets were treated on a different footing, since one
monomer was a helical turn (3.6 amino acids) in the former case or a
single amino acid in the latter. This dissymetry will be corrected
below where we also consider the long range contribution of the
dipole-dipole interaction.

\subsection{The dipolar chain}

Since the peptide bond has a large dipole moment, it seems rather
natural to investigate the properties of the dipolar chain, which
connects successive ${\rm C}_{\alpha}$ carbon atoms. Representing the
peptide bond by a dipole moment is an approximation, and the dipolar
interaction should be modified at short distances. The Hamiltonian of
the model then reads 

\begin{equation}
\label{ham}
\beta{\cal H} = {v_0 \over 2}  \sum_{i \ne j} \delta(\vec r_{i}-\vec
r_{j}) + {\beta \over 2} \sum_{i \ne j} \sum_{\alpha, \gamma}
p_{i}^{\alpha} G_{\alpha \gamma}(\vec r_{i},\vec r_{j}) p_j^{\gamma}
\end{equation}

In equation (\ref{ham}), $\beta={1 \over T}$ is the inverse
temperature, $v_0$ is the excluded volume, $\vec r_{i}$ denotes the
spatial position of monomer $i$ ($i=1,2,...N$), and $\vec p_{i}$ its
dipole moment. If necessary, three body repulsive interactions may be
introduced, to avoid collapse at infinite density. The (infinite
range) dipolar tensor reads 
\begin{equation}
\label{dipo}
 G_{\alpha \gamma}(\vec r,\vec r')= A {1 \over \vert \vec r -\vec
 r'\vert^{3}} (\delta_{\alpha \gamma}-3 v_{\alpha}v_{\gamma})
\end{equation}
with $v_{\alpha}={(\vec r-\vec r')_{\alpha} \over \vert \vec r -\vec
 r'\vert}$ and $A$ is a prefactor containing the dielectric constant
 of the medium.
The dipolar interaction (\ref{dipo}) is modified at small distances
 ($|\vec r-\vec r'|< a$) and may also be cut-off at large distances
 by an exponential prefactor. The partition
 function of the model (\ref{ham}) is given by:
\begin{equation}
\label{partit}
{\cal Z} =\int \prod_{i} d {\vec r_i} d {\vec p_i} \
{\delta(\vert{\vec r_{i+1}-{\vec r_{i}}}\vert-a)} \ \delta (\vert{\vec
p_{i}}\vert -p_0) \  \delta( \vec p_i \cdot (\vec r_{i+1}-\vec r_{i}))
 \ \exp \left( - \beta  
{\cal H} \right)
\end{equation}
In equation (\ref{partit}), $a$ denotes the Kuhn length of the
monomers, and $p_0$ is the magnitude of their dipole moment.  The
third $\delta$-function constraints the dipole moment $\vec p_i$
to be perpendicular to the chain (Figure 1), with full rotation around
bond $(i, i+1)$.

Apart from numerical simulations, there are two main approaches to the
dipolar chain:

(i) one may try to integrate out the dipoles, and get an effective
Hamiltonian for the chain. Since dipoles favor anisotropic configurations,
one expects that the above orthogonality constraint  will lead to
an anisotropic collapsed phase.

(ii) one may tackle the full problem, and using what is known about
ferroelectric domains, one may guess low energy structures.

The latter approach is not without risks, since the determination of domain 
structures from first principles is an unsolved problem in non-soft
condensed matter physics. The former, which is a little more tractable, 
is an extension of the hydrogen bond models (see section \ref
{hbonds}), with one important difference: the long range character of
the dipole interaction, implies that the surface of the collapsed
globule is itself a variational parameter. It also allows for a first
step in accounting for chirality.

\subsection{Integrating the dipoles}
\label{sec: Landau}
{\begin{itemize}
\item Order parameters

For simplicity, we soften the constraint of fixed
length dipoles in equation (\ref{partit}) and replace it by a Gaussian
constraint. We therefore have
\begin{equation}
\label{partit2}
{\cal Z}_G =\int \prod_{i} d {\vec r_i} d {\vec p_i} \
{\delta(\vert{\vec r_{i+1}-{\vec r_{i}}}\vert-a)} \ {1 \over ({2
\pi p_0^2})^{3/2}} \ e^{-{{\vec p_i}^2 \over {2p_0^2}}} \  \delta( \vec p_i \cdot (\vec r_{i+1}-\vec r_{i}))
 \ \exp \left( - \beta  
{\cal H} \right)
\end{equation}
where the subscript {G} on the partition function stands for
Gaussian and the Hamiltonian ${\cal H}$ is given by equation (\ref{ham}). 
Using the identity
\be
\label{identi}
\delta(\vec y)=\lim_{\lambda\to \infty}\left(\lambda \over {2\pi}\right)^{3/2}
e^{-{\lambda{\vec y}^2 \over 2}}
\ee
we may now perform the (Gaussian) integrals over the dipole moments
$\vec p_i$ in equation (\ref{partit2}). As a result, the problem now depends
only on the polymeric degrees of freedom. Introducing
the tensorial parameter $Q_{\alpha\beta}(\vec r)$ by  
\be
\label{paramor}
Q_{\alpha\beta}(\vec r)= \sum_{i}\left({(\vec u_i)_{\alpha}(\vec
u_i)_{\beta}}-\delta_{\alpha \beta}\right) \ \delta(\vec r-\vec r_{i})
\ee
where $(\alpha,\beta=x,y,z)$ and the notation $\vec u_i = (\vec r_{i+1}
- \vec r_i)/a$ was used, we can express the effective Hamiltonian as a
function of the physical order parameters
\be
\label{rho}
\rho(\vec r)= -{1 \over 2} \ {\rm Tr} \ {\bf Q}(\vec r)=\sum_i
\delta(\vec r-\vec r_{i})
\ee
and
\be
\label{q}
q_{\alpha\beta}(\vec r)=\sum_i (u_i^{\alpha} u_i^{\beta}
-{\delta_{\alpha \beta} \over 3}) \ \delta(\vec r-\vec r_{i})
\ee
In a way analogous to the previous section, the density $\rho(\vec r)$ is
appropriate for an isotropic $\theta$ transition, and the dielectric
tensor ${\bf q}(\vec r)$, is appropriate for a (liquid) crystalline
order. Using non rigorous approximations (which are actually valid in
a melt), we find that the dipolar chain undergoes a second order
$\theta$ transition from the coil phase to a (liquid) collapsed phase,
with order parameter $\rho(\vec r)$, followed, at lower temperature,
by a first order transition with order parameter ${\bf q}(\vec
r)$. The ordered phase is a (liquid crystal) collapsed phase. 

\item Trying to include chirality

If one follows liquid crystalline traditions \cite{PG_Pro}, chirality
is usually represented as the simplest non trivial term in a Landau-like
expansion of the free energy in the order parameter ${\bf q}(\vec r)$.
It is well known that chirality is not easy to take into account at a
microscopic level \cite{chiral}, but at a Landau free energy level, symmetry
considerations lead, to lowest order, to a chiral contribution
\be
\label{chiral}
{\cal F}_{chiral}(\vec r)= \int \ D \
\epsilon_{\alpha\mu\nu}q_{\alpha\delta}\partial_{\mu}q_{\nu\delta} \ d^3r
\ee
where $\epsilon_{\alpha\mu\nu}$ the completely antisymmetric tensor
and $D$ is a measure of the strength of the chirality (in fact
several parameters are in general needed).

Using the same non rigorous approximations as above, we find that the
$\theta$ transition is very weakly $D$-dependent, whereas the
transition towards the liquid crystalline phase increases strongly
with $D$ \cite{Pi_Ga_Or}. This liquid crystalline phase is now
modulated, and has strong similarities with the liquid crystalline blue
phase(s). Interestingly enough, for strong enough chirality, we get a
direct transition from a coil phase to a compact phase with a modulated
order parameter ${\bf q}(\vec r)$.
More specifically, one has
\be
\label{blue}
q_{\alpha \beta}(\vec k)=\sum_i e^{-i\vec k \cdot \vec r_i}
(u_i^{\alpha} u_i^{\beta} -{\delta_{\alpha \beta} \over 3})
\ee
in Fourier space (with $\vec k \ne \vec 0$). The indices
$\alpha,\beta$ are space, not replica, indices. The order parameter is 
very similar for the case of (idealized) helices and sheets where

(i) a (ideal) helix is described by $\vec r_i= (u \ {\rm cos} \ v_i, u \
{\rm sin} \ v_i, v_i)$ and constant $u$.

(ii) a (ideal) sheet is described by  $p=1,2,...M$ strands where each
strand is described by a segment $\vec r_{p}=(u \ {\rm cos}
\ v_p, u \ {\rm sin} \ v_p, v_p)$ and constant $v_p$.

To summarize, we have found that within certain models and
approximations, we may get for the dipolar chiral chain, a direct and
discontinous transition from a coil phase to a compact phase with
secondary structures. The order parameter (\ref{blue}) describes both
helix-like and sheet-like conformations.

This model may seem oversimplified. For instance, the chiral
term (\ref{chiral}) can be rewritten as $\sum_{i,j} \left({\vec
\nabla}f(\vec r_{ij}) \cdot (\vec u_i \times \vec u_j)\right) (\vec
u_i \cdot \vec u_j)$, where $f(\vec r_{ij})$ describes a short range
(in space) interaction (the coordinates $\vec r_{i}$  are the
coordinates of the virtual ${\rm C}_{\alpha}$ chain). This term does not
compare well to the dihedral terms of the all atom CHARMM energy which 
read
$$\sum^{ }_{ {\rm dihedrals}}k_\phi( 1+ {\rm cos}(n\phi -\delta))$$

where $\phi(1234) ={\rm cos}^{-1}(\vec n_{123} \cdot \vec n_{234})$
where $\vec n_{abc}$ is the unit normal vector to the plane (a,b,c).
Our modeling of chirality, through a single parameter and a first
order gradient term, is rather primitive. In particular, higher order
gradient terms, describing shorter distances, are certainly important
(see the example of the blue fog in liquid crystalline blue phases)
\cite{Sta_Lub}. Furthermore, the precise spatial organization of the
compact ordered phase also depends on its surface.
\end{itemize}

With all these caveats, it should be mentionned that computing and
diagonalizing ($q_{\alpha \beta}(\vec r)$) for real proteins, leads to
a reasonable characterization of secondary structures. Helices are
essentially uniaxial (two different eigenvalues), whereas sheets are
biaxial (three different eigenvalues) \cite{Delarue}.

\newpage

\subsection{Non integrating the dipoles}

The non integration of dipoles leads one to consider low
energy dipolar structures. I will first recall a few facts on 
ferroelectric domains (structure, order 
parameter,...). More speculative issues, such as 
a ``biological'' interpretation of defects of this order parameter, or
the introduction of surfaces in protein folding, will
be briefly examined.

\begin{itemize}

\item Dipolar ordering

There is another way to consider the dipolar Hamiltonian, namely to
use the identity (see equation (\ref{dipo}))
\be
{(\delta_{\alpha \gamma}-3 v_{\alpha}v_{\gamma}) \over \vert \vec r -\vec
 r'\vert^{3}} ={\partial^2 \over \partial r_{\alpha} \partial
r'_{\gamma}} \left({1 \over \vert \vec r -\vec
 r'\vert}\right)
\ee
Defining a local polarization $ \vec {\rm P}(\vec
r)=\sum_i \vec p_i \delta(\vec r-\vec r_i)$, one may transform the
dipolar term of equation (\ref{ham}) into a Coulomb Hamiltonian, with
a (continuous) distribution of bulk ($\rho(\vec r)=-{\rm div} \ \vec
{\rm P}(\vec r)$) and surface ($\sigma (\vec r_s)=\vec {\rm P}(\vec
r_s)  \cdot \vec {\rm N}(\vec r_s)$) charges,  where 
$\vec r_s$ belongs to the surface, and $\vec {\rm N}(\vec r_s)$ is the
normal to the surface at this point. Given the dimensions and
discreteness of the system, this continuum picture may not be very
satisfactory \cite{Nakamura}, but we will nevertheless use it. 

A (low temperature) collapsed dipolar chain, if long enough, will
break into (Bloch, Weiss, N\'eel...) domains, as I now show on a
simple example. Let us consider an Ising chain, with
short range exchange $(J_0)$ interactions between neighbouring
monomers (see the Appendix  and reference \cite{Orlandini}).
At low temperature, the chain is collapsed and has a uniform
polarization. If one adds a long range dipole-dipole  
interactions $(J_{dd})$, one may test the stability of the uniform
state: flipping half of the dipoles results in an energy cost of order
$\Delta E=+J_0  R^2 -J_{dd} R^3$, where we have considered a spherical
globule of radius $R$, and dropped some numerical constants. For $R
>R^{\star}={J_0 \over J_{dd}}$ (or $N >N^{\star}=({J_0 \over
J_{dd}})^3$), the system will break into domains. 

For many dipolar systems, low energy domain structures in ferroelectricity
(and ferromagnetism) tend to have  $\rho(\vec r)=-{\rm div} \ \vec
{\rm P}(\vec r)=0$ (pole avoidance ``principle''  \cite{Aha} ). 
This ``principle'' is obeyed in the case of helices (where
$\vec {\rm P}(\vec r)$ is a constant vector) and sheets ( where $\vec
{\rm P}(\vec r)$ has roughly antiferroelectric order). One may
understand in a similar way solenoidal
proteins (where $ \vec {\rm P}(\vec r)$ looks like a
curl, i.e. has a circulating pattern) such as $\beta$-barrels or
protein 1a4y (Figure 10). 

\begin{figure}
\begin{center}
\includegraphics[height=7cm]{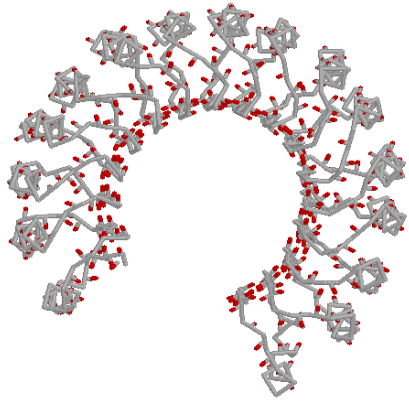}
\caption{Circulating CO dipoles in protein 1a4y}
\end{center}
\end{figure}

The fact that dipolar interactions couple the $\vec r_i$'s and the $\vec
p_i$'s implies that dipolar systems are very sensitive to the
geometry. For instance, the infinite simple cubic and face centered cubic
lattices do not have the same ground state order. For finite systems, the
situation is even more tricky, since dipolar interactions ``feel'' the 
surface of the system. The formation of domains results in general
from the competition between dipolar- and other (shorter range) -
 interactions. The full determination of domain structures (size,
order parameter, spatial organization,...) in non-soft condensed
matter physics depends on non-extensive terms in the free energy, and
remains an unsolved problem \cite{Aha}.

At a smaller (cluster) scale, Singer and coworkers
\cite{Singer1} have studied the ground state of some
dipolar clusters without the chain constraint. They pointed out
that possible order parameters are the vectorial spherical harmonics
(VSH) \cite{Varsha}, and that circulating patterns have then simple
expressions. Writing 

$$\vec {\rm P}(\vec r)=\sum_{(JM)} \sum_{L} A_{JL1}^{M}(r)   \vec
Y_{JL1}^{M}(\theta,\phi)$$ 

one may follow what has been done \cite{Nel_Spa} for (scalar) bond
orientational order in clusters. In this case, one expands the density
as 
$$\rho(\vec r)=\sum_{(LM)} Q_{LM}(r) Y_{LM}(\theta,\phi)$$
where the $Y_{LM}(\theta,\phi)$ are the ordinary spherical
harmonics. The (VSH) formalism is rather heavy, and I have done only
preliminary calculations on real proteins. As expected, helices are
easier to analyze than sheets. 

\item Speculations on order parameters and defects

One may also wonder whether the existence of an order parameter may
help us to understand the existence of an active site, see
e.g. \cite{Foote,Ondrechen}. The simplest idea I can think of is
to view the active site as a defect of this order parameter (see
e.g. \cite{Merm,Klem1}). By defect, I mean here a topological
defect. 

In a first approach, we have integrated out the dipoles and
found an order parameter $q_{\alpha\beta}(\vec r)$, endowed with
complicated line (and other) defects \cite{Merm,Klem1}). I will not
consider these defects here.

If we do not integrate the dipoles, the order parameter is
the local polarization or the local electric field. But real proteins
have also charged residues: since dipoles tend to order along their local
electric field $\vec e$, there seems to be a competition between the
${\rm div} \ \vec {e}=0$ order of the main chain and a ${\rm div} \ \vec
{e} \ne 0$ order of the charged residues. As a result, an active site
would tentatively result. One possibility is to consider the
topological defects of a vector field in three dimensions (namely
points and non singular textures). An interesting  
example is given in ref. \cite{Ripoll} where it was found that half of 
the electric flux through the active site of some $\beta$-barrels came
from the main chain. On the other hand, non singular textures can be
understood with the following example: a (${\rm div} \ \vec {e}=0$)-
ordering implies that one may write $\vec {e}= {\rm curl} \ \vec
c$. One may then calculate $\int \vec {e} \cdot \vec {c} \ d^3r$ over
the volume of the system, and this quantity, if not zero, has
topological meaning \cite{Ranada}. 

Two comments make these speculations even more speculative: to
associate the active site with some defects of an order is a classical 
(not quantum) view. Moreover, the chain constraint has been forgotten:
the equilibrium state of a chain depends also on the chain
conformational degrees of freedom: taking a flexible continuous string
as an example, mechanical equilibrium implies for instance that ${\rm
T}-{\rm V}={\rm Cst}$ along the string, where T is the tension of the
string and V the potential energy per unit length. Any variation (or
defect) in the electrostatic part V will also show up in the
conformational degrees of freedom (represented here by T). It is
interesting to note that a recent paper computes knots-related
invariants with respect to the $\vec r_i$'s degrees of freedom
\cite{Levitt,DeSantis,Fain}. These invariants may have an
electrostatic interpretation \cite{Moffatt}.
\item Possible connections with more geometric approaches

Various geometrical aspects of proteins have been recently stressed,
mostly in relation with packing properties \cite{Maritan,Soyer}. An
older connection concerns surfaces \cite{Salemme,Louie,Hyde,Bahar}
which were argued to be important in protein folding.

From an electrostatic point of view, one could naively expect
positive (resp. negative) charges of a protein to be in a negative
(resp. positive) electrostatic potential. Since the main backbone
chain has both types of charges, it should be close to an
equipotential (not necessarily minimal) surface. The same type of
description applies to hydrophilicities variables ($\lambda_i$), and
suggests a-curve-on-an-interface (or surface) view of a protein. 

Most prominent among these surfaces are minimal surfaces
(i.e. surfaces with zero mean curvature) \cite{Nit}, which have been
introduced in various physical problems, including blue phases
\cite{Pa_Du}. The surface view of proteins is interesting: the ideal
secondary structures described by equation (\ref{blue}) can also be
described as the asymptotic curves of a helicoid. There is ample room
in the geometry of surfaces for the existence of an active ``site''
(focal surfaces, flat points,...), but  more importantly, there are
remarkable surface to surface transformations \cite{Nit}, with very
low energy barriers. Clearly, it would be interesting to implement
these transformations on a computer.

\end{itemize}

\section{Simplified models: heteropolymeric approaches}

There are many recent reviews of the freezing transitions of various
heteropolymeric models see e.g. \cite{Garel,Dill,Shakh,Grosberg,Dev}. I 
will therefore make a sketchy presentation 
of this approach. These models, which are something like spin glasses
\cite{Mezard} with a chain constraint, are difficult to solve, except
in high enough dimensions. The free energy of the frozen phase is
determined, as in spin glasses, by subtle non-extensive terms. One
views a protein as a random polymer with a fixed disordered sequence,
corresponding to the primary sequence. Analytical methods (I will not
consider here numerical simulations) broadly fall into two classes 

(i) replica calculations, where one
averages the disorder over some distribution. 

(ii) calculations where the disorder is not averaged (TAP-like self
consistent field equations, Imry-Ma arguments, variational calculations,
dynamical equations,...).

These models take a rather coarse grained view of the protein, so that  
a residue $R_i$ is represented by a monomer $i$ at position $\vec
r_i$, with some random characteristics (polar character or
hydrophilicity $\lambda_i$,
charge $q_i$,....). The disorder is mostly
included in two body interactions, with various
distributions. Denoting by $r_{ij}$ the
distance between monomers $i$ and $j$, and by $f(r_{ij})$ a short
range interaction, some models of interest are
\begin{itemize}
\item The randomly charged (RC) chain

\be
\label{chargee}
{\cal H}_{RC}= +\sum_{ij}q_iq_j \ f(r_{ij})
\ee

\item The HP chain

\be
\label{hp}
{\cal H}_{HP}= \sum_{ij}(a\lambda_i\lambda_j+b(\lambda_i+\lambda_j)) \
f(r_{ij})
\ee
For the particular value $a=0$, the HP chain is often referred to as the
the random hydrophilic hydrophobic (RHH) chain.

\item The Hopfield (HO) chain

Each monomer $i$ has $M$ generalized charges $(q_i^p; p=1,2,..,M)$, so 
that 
\be
{\cal H}_{HO}= \sum_{p=1}^M \sum_{ij}q_i^pq_j^p \ f_p(r_{ij})
\ee
Its coding properties can be studied in a way analogous to spin
glasses. 
\item the Random Energy Model (REM) chain

It is a kind of $M \to \infty$ limit of the Hopfield chain.
\be
\label{reme}
{\cal H}_{REM}= \sum_{i<j}v_{ij} \ f(r_{ij})
\ee
with uncorrelated values of the {\it pair} interactions $v_{ij}$ and
$v_{kl}$. As in spin glasses \cite{Derrida}, this model is a kind of
fixed point for the physics of heteropolymers \cite{BW}.
\end{itemize}

\vskip 3mm
Note that chirality is seldom included in these models, since it is
not usually expressed as a two body interaction. The phase diagrams of
these models are approximately known in the thermodynamic limit,
in high enough dimensions $d$, and for independent disorder variables
(e.g. $\lambda_i$).

As in the homopolymeric approach, there are (at least) two coil
phases, one above a $\theta$ transition and the other above a freezing
transition (with slow dynamics). In the case of  the (RHH) chain, a
Flory-Imry-Ma approach yields a disorder dependent free energy 
\be
\label{Flory}
F_{RHH}= {R^2 \over N}+ (v_0 +\beta \lambda_0 +u{\beta \lambda \over
\sqrt{N}}){N^2 \over R^d}+w{N^3 \over R^{2d}}
\ee
where $\beta={1 \over T}$, $\lambda_0$ (resp. $\lambda$) is the mean
(resp. variance) of the distribution of hydrophilicities
$(\lambda_i)$, u is a symmetric (Gaussian) random number of variance
one, and $v_0,w$  are two and three body interactions.  
In the coil phase, i.e. at high enough temperature, one may
characterize the short range order by the $u$-dependent
term. For $u>0$ (hydrophilic fluctuation), small $N$ behavior (that is 
for $N< N_0(u) \sim u^2 ({\beta\lambda \over v_0+\beta\lambda_0})^2$)
can be extracted from a Flory estimate ${R^2 \over N} \sim u\beta h
{N^{3 \over 2} \over R^d}$, yielding a branched polymer short range
order \cite{Lu_Is}. On the other hand, a region with $u<0$
(hydrophobic fluctuation) is locally more collapsed (with $R \sim N^{3
\over 2d}$).

Since proteins are not random, it is important to connect real primary
sequences to random ones. A conservative statement is that proteins
have the choice between a quick collapse transition towards the
structureless $\theta$ globule or a slow freezing transition towards a
more structured (frozen) globule. A possibility \cite{Shakh2,Thirum}
is that real sequences 
fold along a kind of Nishimori line \cite{Nishimori} in the phase diagram
(that is a line separating coils with different short range
order). Another possibility is to introduce long range correlations 
(along the chain) in the distribution of the disorder \cite{Koklov}.

Finally, as far as I can see, the existence of an active site is not a major
issue in heteropolymeric models.

\section{Conclusion}

I have discussed some \cite{Domany} homo- and hetero- polymeric aspects of
proteins. The main chain, which refers to the former point, suggests
a connection between folding and ferroelectric (or ferromagnetic) domain
theory, through the model of a chiral dipolar chain. On the other
hand, the side chains point towards a (spin) glass analogy, if their
physico-chemical properties are represented by disorder variables
(charges, hydrophilicities,...). Different length scales related
either to domain formation (e.g. $N^{\star}$) and stability
(e.g. $l_c$), or to disorder fluctuations (e.g. $N_0(u)$), have been
shown to arise in this ``dipolar Imry-Ma'' problem \cite{Nattermann}.

A recent paper \cite{Pado} studies the prediction of bubble and stripe
domains in uniaxial (Ising) ferromagnetic sytems: the 
long range dipolar interaction is relevant only to define the size of the
individual bubbles and one may then treat the physics of the problem
(bubble bubble interaction, bubble to stripe transition,...) through
the use of short range interactions {\it only}.

Following the domain theory appeal leads one to consider the
disorder variables at the length scale of secondary structures, and
not for individual residues. As we have seen, this length scale is of
order 10-20 residues for helices and  5-8 residues for individual
strands. The case of proteins is certainly more complicated than the
ferromagnet (chain constraint, solvent,..), but there have been
important progresses along somewhat similar lines \cite{Baker}.

As for the dynamics of the folding, a very puzzling question remains:
`how does a protein find its way in phase space''? The answer may
require a detailed knowledge of the unfolded phase (short range
order, topological invariants or defects,...). 
\vskip 30 mm

{\bf Acknowledgments:} Most of the work described in these notes has
been done in collaboration with H. Orland. It is a pleasure
to thank J. Bascle, A. Boudaoud, M. Delarue, S. Doniach, S. Franz,
J-R. Garel, M. Kl\'eman, C. Monthus, E. Orlandini, B. Pansu,
P. Pieranski, E. Pitard and P. Rujan for past and future interactions. I am 
grateful to H. Rosenberg for an introduction to minimal surfaces, to
S. Lidin and S. Hyde for sending (p)reprints of their work on minimal
surfaces, and to S. Padovani for explanations and discussions on his
neural network approach to magnetic domains. Last but not least, I
would like to thank Amos Maritan and Jayanth Banavar for their kind
invitation at this meeting.

\newpage

\centerline{\bf Appendix}
\vskip 4mm
\centerline{\bf Some properties of $ {\rm n} \to 0$ spins}
\vskip 4mm
Consider an ${\rm n}$-dimensional classical spin $\vec {\rm S}$ with
\be
\label{A1}
{{{\vec {\rm S}} \ ^{2}=\sum_{u=1}^{{\rm n}} {\rm
S}_{u}^{2}={\rm n}}}
\ee
Defining a {{normalized}} measure $d{\mu}(\vec S)$ on the $({\rm
n}-1)$ dimensional sphere, the average of a function $A({\vec {\rm S}})$
is given by

$$<A({\vec {\rm S}})>=\int d{\mu}(\vec S) \  A({\vec {\rm S}})
$$
\vskip 5mm
An important exemple is the O(n) symmetric function $f(\vec k)=f(k)$
defined by

 $$f(k)=<e^{i\vec k \cdot {\vec {\rm S}}}>$$
One has 
\be
\label{A2}
{\Delta f(k)}=\sum^{\rm n} _{u=1} {\partial ^{2} f \over
\partial k_{u}^{2}}=-<\left(\sum_{u=1}^{{\rm n}} {\rm
S}_{u}^{2}\right) \ e^{i\vec k \cdot {\vec {\rm S}}}>=-{\rm
n}f(k)
\ee

Since $${\Delta f(k)}={d^2 f \over dk^2}+{n-1 \over k}{df \over dk}$$

one finds that ${\rm lim}_{{\rm n}\to 0} \ f(k)=1-{k^2 \over 2}$,
implying the following results:

{{$< {\rm S}_{u}^2>=1$

 $< {\rm S}_{u}^p>=0$ for $p>2$
\vskip 3mm
$<{\rm S}_{{u}_1}^{p_1} {\rm
S}_{{u}_{2}}^{p_2}......{\rm S}_{{u}_r}^{p_r}>=0$}}

which imply
 $$<e^{{\vec {\rm H}} \cdot \vec {\rm S}}>=1+{{\vec {\rm H}}^2 \over 2}$$.
\medskip
\vskip 4mm
\centerline{\bf {Application to Self Avoiding Walks}}
\vskip 4mm
Consider a lattice $(\vec r)$ and define an ${\rm n}$-dimensional
spins ${\vec {\rm S}}_{r}$ at each lattice site. Let us consider the
following quantity:
\be
\label{A3}
{\cal Z}_N=\int d{\mu}(\vec S_{r}){1 \over N!} \ {\left(\sum_{<
rr'>}{\vec {\rm S}}_{r} \Delta_{{\vec r}{\vec r'}}    {\vec {\rm
S}}_{r'}\right)}^{N}
\ee
where $\Delta_{{\vec r}{\vec
r'}} =1 $ if {sites $r$ and $r'$ are 
nearest neighbour, $0$ otherwise (and $\sum_{<rr'>}$ is the sum over
the bonds). 

Using ({$< {\rm S}_{u}^2>=1$ and $< {\rm S}_{u}^p>=0$ for
$p>2$), we see that each vector ${\vec {\rm S}}_{r}$ occurs twice in
${\cal Z}_N$. Since ${\vec {\rm S}}_{r} \cdot {\vec {\rm 
S}}_{r^{\prime}}=\sum_{u=1}^{{\rm n}} {\rm S}_{u r}{\rm
S}_{u r^{\prime}}$, the number ${\cal M}_N$ of
closed Self Avoiding Walks (SAW) of $N$ steps on the lattice is given by
{{$${\cal M}_N={\rm lim}_{{\rm n}\to 0} \ {1 \over {\rm n}} {\cal
Z}_N$$}}.

\vskip 4mm
\medskip
\centerline{\bf More convenient representation of ${\cal M}_N$}
\vskip 4mm
Defining the grand canonical partition function $${\cal
Z}(K)=\sum_{N=0}^{\infty} {\cal Z}_N K^N$$ we get
\be
\label{A4}
{\cal Z}(K)=\int d{\mu}(\vec S_{r}) \ e^{{K \over 2}\sum_{(r,r')}{\vec
{\rm S}}_{r} \Delta_{{\vec r}{\vec r'}}    {\vec {\rm S}}_{r'}}
\ee

where the sum ($\sum_{(r,r')}$) is now over the sites.
The Hubbard-Stratanovich transformation  and the properties of ${\rm
n}$-dimensional spins, as ${\rm n}\to 0$, yield

{{$${\cal Z}(K)=\int {\cal D} {\vec {\varphi}_r} \ e^{-{1 \over 2}\sum_{(r,r')}{\vec
{\varphi}}_{r} \ (\Delta^{-1})_{{\vec r}{\vec r'}} \ {\vec
{\varphi}}_{r'}} \prod_{r}\left(1+{K \over 2}{\vec
{\varphi}_r}^2\right)$$}}

Since ${\cal Z}_N$ is the coefficient of $K^N$ in ${\cal Z}(K)$, we
finally have

$${\cal M}_N={\rm {lim}}_{{\rm n}\to 0}{1 \over {\rm n}}K^{-N}{\cal
Z}(K)$$ 

\begin{itemize}
\item {\bf {Fully compact SAW}}
\vskip 3mm
Fully compact SAW's (i.e. Hamiltonian Paths) are obtained by taking 
the monomer fugacity $K\to \infty$. We therefore get

$${\cal N}_N={\rm lim}_{{\rm n}\to 0} {1 \over {\rm n}} \int
{\cal D}{\vec {\varphi}_r} \ e^{-{1 \over 2}\sum_{(r,r')}{\vec 
{\varphi}}_{r} \ (\Delta^{-1})_{{\vec r}{\vec r'}} \ {\vec
{\varphi}}_{r'}} \prod_{r}\left({1 \over 2}{\vec
{\varphi}_r}^2\right)$$
\be
\label{A6}
\ee

A homogeneous saddle point on ${\vec {\varphi}_r} $ gives
$${\cal N}_N=\left({q \over e}\right)^N$$ where $q=2d$ is the lattice
coordination number. Alternatively, we could have avoided the grand
canonical approach and establish the equality of the two members of
equation (\ref{A6}) through the use of Wick's identity
$$ \overline{ \varphi^{u} \left(\vec  r
\right)\cdot \varphi^{v} \left(\vec  r^{\ \prime} \right)} =
\delta_{uv} \Delta_{\vec  r\vec  r^{\
\prime}}$$
with $(u,v=1,2,...n)$.

\item {\bf Fully compact ``helices''}:

Introducing a curvature energy to disfavor corners in the Hamiltonian
Paths, the partition function reads

\vskip 2mm
$${\cal N}_h=\sum_{(HP, corners)} e^{-\beta N_{corners}(HP)\varepsilon_h}$$

Let us now introduce, for each site
$(r)$ of the lattice,  $d$  ${\rm n}$-dimensional fields: ${\vec
{\varphi}}_{\alpha r}$ ($\alpha=1,2,...d$). Generalizing equation
(\ref{A6}), we now obtain equation (\ref{NHP2}):

$${\cal N}_h={\rm lim}_{{\rm n}\to 0} {1 \over {\rm n}} \int {\cal
D}{\vec 
{\varphi}}_{\alpha r} \ e^{-{1 \over 2}\sum_{\alpha=1}^{d} \sum_{(r,r')}{\vec 
{\varphi}}_{\alpha r} \ (\Delta_{\alpha}^{-1})_{{\vec r}{\vec r'}} \ {\vec
{\varphi}}_{\alpha r'}}$$
$$\prod_{r}\left({1 \over 2}\sum_{\alpha=1}^{d}{\vec
{\varphi}}_{\alpha r}^2+ e^{-\beta \varepsilon_h}
\sum_{\alpha<\gamma}{\vec {\varphi}}_{\alpha r} \cdot {\vec
{\varphi}}_{\gamma r} \right)$$ 

The identity between the two expressions of ${\cal N}_h$ rely on
Wick's theorem
$$  \overline {\varphi^{ (u)}_\alpha \left(\vec  r
\right)\cdot \varphi^{ (v)}_\gamma \left(\vec  r^{\ \prime} \right)} =
\delta_{ uv}\delta_{ \alpha \gamma} \Delta^ \alpha_{\vec  r\vec  r^{\
\prime}}$$

Performing a homogeneous and isotropic saddle point in equation
(\ref{NHP2}), we get equation (\ref{HEL})
$${\cal N}_h=\left({q(\beta) \over e} \right)^N$$
with an effective coordination number $q(\beta)=2+2(d-1) e^{-\beta
\varepsilon_h}$.

\item {\bf Fully compact ``sheets''}:

Denoting by $N_{bonds}(HP)$ the number of H-bonds in a given (HP), the
partition function reads

$${\cal N}_s = \sum_{(HP, bonds)}{\rm e}^{+\beta \varepsilon_s \
N_{bonds}(HP)}$$ 
where the summation runs over all possible (HP)'s and over
all possible sets of $ {\rm H} $-bonds compatible with this path. 

The formalism is more complicated since the integral representation of
${\cal N}_s$ requires, for each direction $\alpha$:

(i) a ${\rm n}$-component field $ \vec  \varphi_ \alpha \left(\vec
r \right) $ to generate the $(HP)$, with $n \to 0$.

(ii) two scalar fields $ \psi^ +_\alpha \left(\vec  r \right) $ and
$ \psi_ \alpha \left(\vec  r \right) $ which respectively initiate and 
terminate and $ {\rm H} $-bond at site $ \vec  r $ in direction $
\alpha$.

We also have

$${\cal N}_s={\rm lim}_{{\rm n}\to 0} {1 \over {\rm n}}{\int d{\vec 
{\varphi}}_{\alpha}d{{\psi}}_{\alpha}d{{\psi}^{+}}_{\alpha}
 \ e^{-A_G} \prod^{ }_{r} D( r)  \over \int d{\vec 
{\varphi}}_{\alpha}d{{\psi}}_{\alpha}d{{\psi}^{+}}_{\alpha}{\rm e}^{-A_G}}$$
where the (normalizing) denominator is due to the introduction of the
two scalar fields and where
$$ A_G = \sum \left[{1
\over 2}\vec  \varphi_ {\alpha r} \left(\Delta^
\alpha_{r r^{\ \prime}} \right)^{-1}\vec  \varphi_{\alpha r^{\ \prime}}
+\psi^ +_{\alpha r}
\left(\Delta^{ \alpha {+}}_{\vec  r\vec  r^{\ \prime}}
\right)^{-1}\psi_{\alpha r^{\ \prime}} \right] $$

with
$$ D \left(\vec  r \right) = \sum^{ }_ \alpha{ 1 \over 2}\vec  \varphi^
2_\alpha \left(\vec  r \right)\ G_\alpha \left(\vec  r \right) + \sum^{ }_{
\alpha <\delta}\vec  \varphi_ \alpha \left(\vec  r \right)\cdot\vec  \varphi_
\delta \left(\vec  r \right) $$

and
$$ G_\alpha \left(\vec  r \right) = 1 + {\rm e}^{\beta \varepsilon_s /2} \sum^{
}_{ \gamma (\not= \alpha )} \left(\psi^ +_{\gamma r} +\psi_
{\gamma r} \right)+ {\rm e}^{\beta \varepsilon_s} \sum_{
\gamma (\not= \alpha )}\psi^ +_{\gamma r} \psi_{\gamma
 r} $$

The operator $ \Delta^ \alpha_{\vec  r\vec  r^{\ \prime}} $ has the
same meaning as above, and $ \Delta^{ \alpha {+}}_{\vec  r\vec  r^{\
\prime}} $ is 1 iff $\vec r^{\ \prime}=\vec r+\vec e_{\alpha}$, $\vec
e_{\alpha}$ being the unit vector in direction $\alpha$. The two
expressions of  ${\cal N}_s$ can be identified through Wick's theorem:

$$  \overline {\varphi^{ (u)}_\alpha \left(\vec  r
\right)\cdot \varphi^{ (v)}_\gamma \left(\vec  r^{\ \prime} \right)} =
\delta_{ uv}\delta_{ \alpha \gamma} \Delta^ \alpha_{\vec  r\vec  r^{\
\prime}}$$

$$\overline {\psi^{+}_{\alpha r} \psi_{\beta r'}}=\delta_{\alpha
\beta}\Delta^{ \alpha {+}}_{\vec  r\vec  r^{\ \prime}}$$

$$\overline {\psi_{\alpha r} \psi_{\beta r'}}=\overline {\psi^{+}_{\alpha
r} \psi^{+}_{\beta r'}}= 0$$

Performing a homogeneous and isotropic saddle point on the fields ($
\vec  \varphi_ \alpha \left(\vec r \right), \psi^ +_\alpha \left(\vec
r \right), \psi_ \alpha \left(\vec  r \right)$) leads to a
crystallization transition similar to the case of helices.

\end{itemize}

\vskip 4mm
\medskip
\centerline{\bf The Ising chain}
\vskip 4mm
\be
\label{Z}
Z= \sum_{\rm SAW} \sum_{S_i=\pm 1} \exp\left({{\beta J \over 2} \sum_{i \ne j}
S_i \Delta_{r_i r_j} S_j }\right)
\ee
where $J$ is the exchange energy. The sums run over all possible SAW
and all spin configurations. By using a Gaussian transform, 
it is possible to rewrite (\ref{Z}) as
\be
\label{field1}
Z= 2 ^ N \int \prod_{r} d \varphi_r  \exp \left ( -{1 \over 2 \beta J}
\sum_{\{r,r'\}} \varphi_r \Delta ^{-1} _{r,r'} \varphi_{r'} + \log
{\sum_{{\rm SAW} \{r_i\}}
\ \prod_{i=1}^N \cosh ( \varphi_{r_i} )} \right )
\ee
Mean-field theory can be obtained by performing a
saddle-point approximation on equation (\ref{field1}). We assume that the chain
is confined in a volume $V$ with a monomer density $\rho={N\over V}$.
Assuming a translationally invariant field $\varphi$, the mean field
free energy per monomer is
\be
\label{meanfield1}
f = {F \over N} =  -T \log2 + {T^2 \over 2 \rho J q} \varphi^2 -T 
\log Z_{\rm SAW} -T \log \cosh (\varphi)
\ee
where $q=2d$ and $Z_{\rm SAW}$ is the total number of SAW of $N$ monomers
confined in a volume $V$. It is easily seen that
\be
Z_{\rm SAW} \simeq \left ( {q \over e} \right )^N \exp {\big( - V (1-\rho)
\log (1-\rho)\big)}
\ee
so that
\be
\label{meanfree}
f = -T \log2 + {T^2 \over 2 \rho J q} \varphi^2 -T 
\log {q \over e} + T\ {1-\rho \over \rho}\ \log ( 1-\rho)
-T \log \cosh (\varphi)
\ee
This free energy is to be minimized with respect to $\varphi$ and
$\rho$, yielding a discontinuous transition to a compact
ordered phase \cite{Orlandini}.

\end{document}